# The Geometry of Soil Crack Networks

V.Y. Chertkov*

*Agricultural Engineering Division, Faculty of Civil and Environmental Engineering, Technion, Haifa 32000, Israel*

**Abstract:** The subject of this work is the modification and specification of an approach to detail the estimation of soil crack network characteristics. The modification aims at accounting for the corrected soil crack volume based on the corrected shrinkage geometry factor compared to known estimates of crack volume and shrinkage geometry factor. The mode of the correction relies on recent results of the soil reference shrinkage curve. The main exposition follows the preliminary brief review of available approaches to dealing with the geometry of soil crack networks and gives a preliminary brief summary of the approach to be modified and specified. To validate and illustrate the modified approach the latter is used in the analysis of available data on soil cracking in a lysimeter.

## INTRODUCTION

Predicting the geometry of shrinkage crack networks is obviously of crucial importance for modeling water flow and solute transport in soils. This work deals with the essential improvement of an approach [1-5] to detail predicting characteristics of a three-dimensional soil crack network. The improvement is based on the recent results of estimating the corrected crack volume [6, 7] and effects of an intraaggregate structure on the reference shrinkage curve [8-10] of the soil. Preliminarily, we briefly review available approaches to the issue of soil crack network characteristics. First, note that nowadays the problem is not solved by fracture mechanics methods (e.g., [11]) although one includes different criterions of crack development. In particular, for cracks in elastic-plastic soils the Irwin-Orowan criterion (e.g., [5]) or the criterion of the crack-tip opening angle (e.g., [12]) is used. At the present, however, fracture mechanics only operates with a separate crack, or several cracks, or a crack system of some special symmetry (e.g., [13, 14]), and irrespective of a concrete fracture criterion, is unable to regard real multiple cracking with crack network formation because of its essentially statistical nature.

There are a number of approaches which only deal with two-dimensional (2D) crack networks in a thin layer. First, we note 2D macroscopic approaches. Yoshida and Adachi [15] relied on Biot's consolidation theory [16]. These authors only predict the potential location of cracks in an early stage of desiccation. Karalis [17] applied a thermodynamic approach to estimate mean crack spacing only. Perrier et al. [18] suggested possible types of crack networks (including shrinkage-originated ones) based on a fractal model [19]. Moran and McBratney [20] simulated crack networks using a linked distribution of points. Horgan and Young [21] considered a model generating crack networks from crack growth development as a random walk. It is worth noting that the direction of generating a final crack network can also be opposite to that in the three last mathematical approaches. Indeed, they first simulate a network of the largest cracks (macrocracks) and fragments. Then, these fragments are covered by a secondary crack network and so on. In natural materials such as soils and rocks, in natural processes leading to shrinkage such as drying and cooling, cracking and crack network formation can go in reverse direction starting from microcrack appearance to accumulation, coalescence, and development.

There is also a 2D microscopic approach. It is presented by a number of physical models that simulate a 2D crack network starting of a microlevel (e.g., [22-26]). All these models are essentially very close each other and replace a thin shrinking continuous layer by a network of horizontal elastic springs that connect discrete nodes. Every node is linked by the springs with six neighbors. At the bottom the nodes either slip with friction [24, 26] or are attached by vertical springs. The accumulation of broken springs at shrinkage (according to a simple fracture criterion) leads to a crack origination and development up to a crack network formation. These models are obviously oversimplified compared to reality even for the 2D case because at the deformation of cohesive soils they only keep the effects of elasticity and ignore the contributions of inertia (the motion of masses in nodes), viscosity, and plasticity. There is no clear indication what particles (masses) are implied in the nodes (clay particles, clay aggregates, silt-sand grains, or something else); that is, what is understood under a microscale which determines the size of primary cracks (microcracks).

---

*Address correspondence to this author at the Agricultural Engineering Division, Faculty of Civil and Environmental Engineering, Technion, Haifa 32000, Israel; E-mail: agvictor@tx.technion.ac.il

Thus, it is questionable that all three above mathematical 2D macromodels and the noted physical 2D micromodels, can imitate a real process of 2D crack network formation in real natural materials. At the same time, all these different models show a qualitative resemblance between the simulated and observed 2D crack networks, and this is a major argument of their feasibility. This contradiction between the obviously rough simulation (for all these different models) and qualitative similarity of simulated and observed networks in all the cases should be explained by some general reason. The above contradiction can be explained, at least in part, based on the intersecting surfaces approach (ISA) to soil structure [27, 28]. The gist of the approach is as follows. The large number of intersecting surfaces (S-objects, e.g., cracks) in a volume divides the latter into sub-volumes (V-objects, e.g., fragments). Then, a certain universal size distribution of the outlined V-objects originates from a number of simple geometrical conditions imposed on the S-objects. In the simplest case the elementary ISA distribution of V-objects is a universal function of the relative V-object size, $x/x_m$ ($x_m$ is the maximum size of V-objects). This means that the view of the function does not depend on S- and V-object type (e.g., soil cracks and fragments or clay particles and clay matrix pores) and methods of their preparation or simulation (e.g., from the above mathematical macromodels or physical micromodels). Materials and methods can only influence the values of the distribution parameters. In the simplest case these are $x_m$ and a connectedness, $c$ of S-objects (e.g., cracks), that is, a ratio of connected to the total number of cracks (or $P$ being the volume fraction of V-objects of all initial volume; $P$ is coupled with connectedness). The ISA approach is also applicable to the 2D case [29].

Thus, any simulated crack network (an intersecting surfaces system) and observed crack network (another intersecting surfaces system) in any case introduce the same (universal) fragment size distribution. That is, simulated and observed crack networks in any case can only differ (in the simplest case) by the scale parameter, $x_m$ and crack network connectedness. Outwardly, these networks can appear to be similar. For this reason a formal qualitative resemblance between a model simulated crack network and an observed one cannot serve as a criterion or proof that a model mimics a real process of crack network formation.

There are also a number of three-dimensional approaches to soil volume cracking. The known macro-approach is based on a close link between variations of the matrix volume, crack volume, and thickness of a soil layer at shrinkage [30, 31]. This link is realized through the concept of the shrinkage geometry factor of a soil layer, $r_s$, which is determined by the change of any pair of the above three values, e.g., the layer thickness and matrix volume. Conversely, known $r_s$, together with one of the above three values, determine two others. However, the approach has a number of drawbacks. The first is that the total crack volume in a layer is only estimated; the distributions of the crack network parameters are not. The second is connected with the determination of $r_s$ (that relates to a soil layer) from core sample measurements [31]. As recently shown, the $r_s$ value found in this way can contain an essential inaccuracy which grows with drying [6, 7]. The inaccuracy is connected with (i) the $r_s$ dependence on water content, (ii) possible crack development inside the core sample, and (iii) presentation of a real layer by a set of disconnected samples. The third drawback is that the latter presentation of a real soil layer only implies the existence of sub-vertical cracks. At the same time the total crack volume includes both the sub-vertical and sub-horizontal cracks. There is some development of Bronswijk's approach [32, 33]. This development suggests the additional calculation of the total vertical crack cross-section area in a layer (at a given depth or water content) using the shrinkage curve. Note that the calculation inherits all the drawbacks from Bronswijk's approach.

A brief summary of the approach [1-5] that we intend to essentially improve is given in the following section. For this reason we only mention its most general points here. The approach combines micro- and macroscopic concepts. A concentration criterion of crack accumulation and merging, starting from microcracks [34], and the effective independence of cracks in the case of multiple cracking [35, 36] are a basis for the modeling of a crack network. This basis enables one to introduce a condition of fragment formation at crack connection and a number of relevant concepts (crack connection probability of the $x$ dimension, fragment formation probability, average and maximum fragment sizes, and crack connectedness) as well as to suggest quantitative relations between the concepts. In the frame of an application the maximum fragment dimension and crack connectedness can depend on the spatial coordinates and parameters specific for the application [27, 28]. One of the applications relates to the shrinkage crack network geometry in swelling soils; the spatial coordinate being the soil depth; the specific parameter being the ratio of an upper layer thickness of intensive cracking to the maximum crack depth. It is assumed that after formation of the vertical crack network thin layers of drying soil along the vertical-crack walls tend to contract, but the moist soil matrix hinders this. This causes the development of horizontal cracks or close to them starting from the walls of vertical cracks [2, 3]. Unlike the above approaches, the probabilistic model under consideration is capable of giving a detailed estimate of crack network geometry at soil shrinkage, namely, the distribution of any crack characteristic at a given soil depth (crack spacing, volume, cross-sectional area, width, depth, cross-sectional trace length) and evolution of the distribution with depth (at a given water content profile). Recently, the usefulness of the approach for applications in hydrology [37, 38] and soil structure [39] has been confirmed. However, the approach has the following major drawback. For the calculation of the distributions of the vertical and horizontal crack characteristics in the frames of the approach, the total crack volume in a layer is needed (see the following section). This total crack volume was calculated using the $r_s$ concept from Bronswijk [30, 31] with all the above mentioned inaccuracies (see below in more detail). The major objective of this work is to remove this drawback based on the recent relevant results [6-9] and show a particular example of analysis of available experimental data using a modified approach. In addition, some approximations (see below) that were used in [1-3] in estimating the characteristics of vertical and horizontal cracks should be specified. This is an additional objective of this work. Notation is summarized at the end of the paper.

Two final remarks concerning the above brief review can be useful. First, instead of the division between the above approaches to the shrinkage crack network in soils as two-, three-dimensional or micro-, macroscopic ones, one can classify the approaches as geometrical [18, 20, 21], dynamic [15, 17, 22-26, 30-33], and mixed [1-5] ones. Second, in the dynamic and mixed approaches the dynamics is introduced in two different ways using either forces induced by hypothetic elastic springs [22-26] or a really observed soil shrinkage curve in a different form [1-5, 15, 17, 30, 33].

# THEORY

**Brief Summary of the Approach under Consideration**

*Three-Dimensional Crack Network with Negligible Crack Volume in Homogeneous Conditions*

The concepts of multiple cracking and fragmentation underlying Chertkov and Ravina's model [1] include presentation of any macrocrack as the result of a number of random sequential coalescences of increasingly larger cracks across a range of spatial scales, beginning from microcracks. Then, the 3D crack network developing in a volume and leading to its fragmentation in statistically homogeneous conditions, is described by the probability $f(x)$ of the connection of cracks of any orientation of dimension $<x$ (or volume fraction of fragments of all the dimensions $<x$), and the volume fraction occupied by all the fragments, $f_m = f(x_m)$, or the fragment formation probability [$(1-f_m)$ is the portion of non-fractionated volume] as

$f(x)=1-\exp[-I(x)]$, (1a)

$I(x)=\ln(6)\ C\ (x/d)^4 \exp(-x/d),\qquad 0<x<x_m$ (1b)

and

$f_m=1-\exp(-8.4C)$ (2)

where for rocks and soils $0<C<1$ is crack connectedness; $d$ is average crack spacing; and $x_m=4d$ is the maximum dimension of fragments. The above concepts, as well as Eq.(1) and (2), allow one to find the network itself, that is, distributions of crack spacing and crack dimensions in horizontal or vertical cross-section, but not distributions for crack width, cross-section area, and volume, that are determined by other considerations (see below). In this approximation we deal with the network with negligible crack volume.

*Three-Dimensional Crack Network with Negligible Crack Volume in Conditions of a Vertical Soil Water Content Profile*

In the indicated inhomogeneous conditions two characteristic soil depths ($z_m$ and $z_o$) are introduced to determine the crack network geometry in horizontal and vertical cross-sections (this means, distribution of crack spacing and so on, but, again, not crack width and volume), that is, depth dependencies, $d(z)$, $f_m(z)$, and $c(z)$ [1, 4]. $z_m$ is the maximum crack depth. $z_o$ is the thickness of an upper layer (a few tens of centimeters) of intensive cracking. By definition of $z_o$, $d(z_o)=z_o$. At the initial transitional stage of drying the concept of intensive cracking is not feasible. That is, this condition cannot be satisfied and for some time the layer simply does not exist. After that when the condition has meaning $z_o$ varies in the range $0.1<z_o/z_m<0.2$. Such a crack system is referred to as a quasi-steady one. In real soils one practically always deals with this stage of crack system development. The case of $z_o/z_m \cong 0.1$ is of practical interest. Typically, a crack network changes with time so slowly that one can consider the situation at a given moment as steady. That is, $d(z)$ does not explicitly depend on time, but only through parameters $z_o$ and $z_m$. $z_m$ can be estimated by the depth of the ground water level. Then, $z_o \cong 0.1 z_m$. Except for that $z_o$ can be estimated as $z_o \cong s$ (then $z_m \cong 10\ z_o$) where $s$ is the simply measurable mean crack spacing at the surface. Note, that at the soil surface the number of small cracks is very large ($d(0) \cong 0$), and the measurable mean crack spacing, $s$ accounts for cracks only with width (at the surface) more than the diameter of a flexible wire (~1.5 mm) for crack depth measurement. Finally, $s$ can be estimated using a number of measurable fundamental soil properties [5].

Two characteristic depths ($z_m$ and $z_o$) determine crack network evolution with soil depth in conditions of the vertical soil water content profile, through network parameters $d$, $f_m$, and $C$ as

$$d(z) = z_o(z/z_o)^\omega, \qquad \omega = 1 - 2\ln 2/\ln(0.8 z_m/z_o), \tag{3}$$

$$f_m(z) = \{1 + \exp[(z - 0.8 z_m)/z_o]\}^{-1}, \tag{4}$$

$$C(z) = \ln\{1 + \exp[-(z - 0.8 z_m)/z_o]\}/8.4 \tag{5}$$

as well as the (total) specific length, $L$ of vertical crack traces in a horizontal cross-section at depth $z$ (per unit area) as

$$L(z) = \{1 - [1 + z_m/d(z)]\exp[-z_m/d(z)]\}/d(z). \tag{6}$$

These dependencies determine the distributions of crack spacing and crack dimensions in horizontal and vertical cross-sections for any $z$ in the above conditions inhomogeneous along the soil depth axis.

### *Three-Dimensional Crack Network with Finite Crack Volume in Conditions of a Vertical Soil Water Content Profile*

Distributions of crack width, cross-section area, and volume at a given depth and their evolution with soil depth are determined by shrinkage in different forms (in addition to characteristic soil depths, $z_m$ and $z_o$ that determine through Eq.(1)-(6) the crack network itself when cracks are considered to be mathematical surfaces without width and volume). First, we consider the vertical crack system [1] because horizontal cracks originate from the verticals (see below).

Denoting $A_o$ and $A(z)$ as the initial and current total area of a horizontal uncracked soil cross-section at a depth $z$, respectively, and $\Delta A(z) \equiv A_o - A(z)$ as an increment of the uncracked area under shrinkage at a given depth $z$, one may introduce the specific horizontal surface shrinkage per initial unit area (or the total specific crack cross-section area at depth $z$), $\delta_A(z) \equiv \Delta A(z)/A_o$. Then, the mean width, $R(z, h)$ at depth $z$ of cracks with tips at depth $h$ ($z<h<z_m$) is

$$R(z, h) = \int_h^z d\delta_A(z')/L(z'). \tag{7}$$

Using for $\delta_A(z)$ the approximation as

$$\delta_A(z) = \int_{z_m}^z R(z, h) dL(h), \tag{8}$$

the term $\pi(z, h)$,

$$\pi(z, h) = -R(z, h) \frac{dL(h)}{dh} / \delta_A(z) \tag{9}$$

is the differential fraction (probability density) of the total specific crack volume (per unit volume of soil), or of the total specific crack cross-sectional area (per unit cross-sectional area), at a depth $z$ related to cracks with tips in a unit interval at depth $h$ ($z<h<z_m$). Furthermore, $P(z, h)$,

$$P(z, h) = \int_h^{z_m} \pi(z, h') dh' \tag{10}$$

is the cumulative fraction of the total specific volume of vertical cracks, or of the total specific cross-sectional area of vertical cracks at a depth $z$, related to cracks with tips at depths $>h$ ($z<h<z_m$). Replacing in Eq.(10) $h$ by $h(z, R)$ where $h(z, R)$ is a solution of the equation $R = R(z, h)$ (see Eq.(7)), one obtains the cumulative fraction of the total specific crack volume or the total specific crack cross-sectional area at a depth $z$ related to cracks with width $>R$ ($0 \le R \le R(z, z_m)$). Probability density, $\pi(z, h)$ (Eq.(9)) and cumulative probability, $P(z, h)$ (Eq.(10)) determine contributions of the vertical cracks of any type (according to their width or positions of their tips) to the crack volume of a soil layer $z_1 \leq z \leq z_2$ (per its unit area). Note that all the above distributions and contributions are determined by the shrinkage characteristic in the form of the specific horizontal surface shrinkage, $\delta_A(z)$ (Eq.(7) and (9)).

Horizontal cracks, their width and volume appear after formation of the verticals due to the difference in water content between soil matrix and thin drying layers along the walls of the vertical cracks [2, 3]. The model also assumes that, on average, the distribution of the volume and width of horizontal cracks is similar for any vertical profile. Denoting $T_o$ and $T(z)$ as the initial and current thickness of a horizontal soil layer around a depth $z$, respectively, and $\Delta T(z) \equiv T_o - T(z)$ as an increment of the layer thickness under shrinkage and cracking around a given depth $z$, one may introduce the specific linear vertical shrinkage of the soil (per initial unit layer thickness) at the depth $z$, $\delta_T(z) \equiv \Delta T(z)/T_o$. The linear vertical shrinkage, $\delta_T(z)$ determines the subsidence, $S(z)$ of the drying soil along a vertical elevation not containing a vertical crack as a function of soil depth, $z$ as

$$S(z) = \int_z^{z_m} \delta_T(z') dz' . \qquad (11)$$

In particular, $S(0)$ is the subsidence of the soil surface. Unlike the linear vertical shrinkage $\delta_T(z)$, the linear vertical shrinkage at a point on the (additionally drying and shrinking) wall of a vertical crack, $\delta_{cr}(z, h)$ depends on the crack tip depth, $h$, and the depth of the point on the wall, $z \leq h$. At the same depth $z$ ($\leq h$) it is usually $\delta_{cr}(z,h) \geq \delta_T(z)$. Below the crack tip depth (when $z > h$) $\delta_{cr}(z,h) = \delta_T(z)$. The value $\Delta S(z, h)$ as

$$\Delta S(z, h) = \begin{cases} \int_z^h [\delta_{cr}(z', h) - \delta_T(z')] dz', & z \leq h \leq z_m \\ 0, & h < z \leq z_m \end{cases} \qquad (12)$$

is defined as the *potential relative subsidence* at depth $z$ of a vertical profile containing a vertical crack of depth $h$. For $\delta_{cr}(z,h)$ the following approximation is used

$$\delta_{cr}(z, h) = \begin{cases} \delta_T(0), & 0 \leq z \leq R(0, h) \\ \delta_T\left(\dfrac{h[z - R(0, h)]}{[h - R(0, h)]}\right), & R(0, h) \leq z \leq h \end{cases} \qquad (13)$$

(for $R(0, h)$ see Eq.(7)). Using the probability density, $\pi(z, h)$ (Eq.(8)) to average the value $\Delta S(z, h)$ at depth $z$ on all depths $h$ of vertical cracks, $z \leq h \leq z_m$ one can define the *mean potential relative subsidence* (MPRS), $\overline{\Delta S}(z)$ as

$$\overline{\Delta S}(z) = -\frac{1}{\delta_A(z)} \int_z^{z_m} \Delta S(z, h) R(z, h) dL(h) . \qquad (14)$$

Considering all vertical profiles to be similar, the total specific width of the horizontal cracks (i.e., the horizontal ruptures on the walls of vertical cracks) per unit height of a vertical profile or the total specific volume of the horizontal cracks, $v_{h\,cr}(z)$ is

$$v_{h\,cr}(z) = \begin{cases} -\dfrac{d\overline{\Delta S}(z)}{dz}(1 - \delta_A(z)), & \text{if } \dfrac{d\overline{\Delta S}(z)}{dz} < 0 \\ 0, & \text{if } \dfrac{d\overline{\Delta S}(z)}{dz} \geq 0 \end{cases} . \qquad (15)$$

The multiplier $(1-\delta_A(z))$ excludes from the total specific volume $v_{h\,cr}(z)$ of the horizontal cracks at depth $z$ a volume at their intersections with vertical cracks which is already included in the volume of the latter. Replacement of $(1-\delta_A(z))$ with $d(z)$ (Eq.(3)) in Eq.(15) gives an expression for the mean width of the horizontal cracks at depth $z$. Note that the horizontal crack system is determined by both the linear vertical shrinkage, $\delta_T(z)$ and horizontal surface shrinkage $\delta_A(z)$.

6*Relations between Different Characteristics of Soil Shrinkage that Were Used in the Summarized Approach*

Denoting $V_o$ and $V(z)$ as the initial and current volume of a soil matrix (without cracks) in a horizontal soil layer around a depth $z$, respectively, and $\Delta V(z) \equiv V_o - V(z)$ as an increment of the soil matrix volume under shrinkage and cracking around a given depth $z$, one may introduce the specific volume shrinkage of a soil matrix, $\delta_V(z) \equiv \Delta V(z)/V_o$. Note that dependencies $T(z)$, $A(z)$, $V(z)$, $\Delta T(z)$, $\Delta A(z)$, $\Delta V(z)$, $\delta_T(z)$, $\delta_A(z)$, and $\delta_V(z)$ can be written as $T(W)$ and so on with $W=W(z)$ where $W$ is the gravimetric water content and $W(z)$ is a current soil water content profile (i.e., for a given drying duration).

The summarized approach used the known relation [30] as

$$[1 - \Delta V(W)/V_o] = [1 - \Delta T(W)/T_o]^{r_s} \tag{16}$$

with $r_s=3$ at any water content [31]. Then, accounting for the above definitions of $\delta_T(W)$ and $\delta_V(W)$ Eq.(16) gives

$$\delta_T(W) = 1 - [1 - \delta_V(W)]^{1/3} . \tag{17}$$

According to definitions of $T(W)$, $A(W)$, and $V(W)$

$$1 - \delta_A(W) = A/A_o = (V/T)/(V_o/T_o) = (V/V_o)/(T/T_o) = [1-\delta_V(W)]/[1-\delta_T(W)] . \tag{18}$$

Then the replacement of $\delta_T(W)$ in Eq.(18) from Eq.(17) gives

$$\delta_A(W) = 1 - [1 - \delta_V(W)]^{2/3} . \tag{19}$$

In this case $\delta_A(W)$ and $\delta_T(W)$ are connected as

$$\delta_A(W) = 1 - [1 - \delta_T(W)]^2 = \delta_T(W)[2 - \delta_T(W)] . \tag{20}$$

Thus, the summarized approach used the expressions from Eq.(17) and (19) to present the linear vertical shrinkage, $\delta_T(W)$ and horizontal surface shrinkage, $\delta_A(W)$, respectively, through volume shrinkage, $\delta_V(W)$. In turn, $\delta_V(W)$ was found from the usual soil shrinkage curve, $V(W)$ (specific soil volume vs. gravimetric water content) ordinarily measured on core samples (e.g., [33]). Finally, validating the approach, Chertkov and Ravina [1] only compared the predicted volume of vertical cracks with measured crack volume.

In general, the relations given by Eq.(17), (19), and (20) should be modified because the $r_s=3$ value should be replaced with an $r_s(W)$ dependence. In addition, finding of the $r_s(W)$ dependence itself should be modified compared to Bronswijk [31]. Finally, a predicted crack volume should include the contributions of both the vertical and horizontal cracks. These modifications as well as some specifications of the approach under consideration and their consequences are discussed and illustrated in the following part of the work.

**Modification and Specification of the Approach to Soil Crack Network Geometry**

*Calculation of the Corrected Shrinkage Geometry Factor and Corresponding Linear and Surface Shrinkage*

Recently, an analysis [6, 7] showed that Bronswijk's approach [30, 31] to estimating the total crack volume in a soil layer (field conditions) through the shrinkage geometry factor, $r_s$ relies on implicit assumptions that are violated in real conditions. First of all, the $r_s$ factor is experimentally estimated by measurements of initial volume and volume as well as the subsidence of cylindrical soil samples after oven drying [31]. That is, the $r_s$ factor is assumed to not depend on water content (usually the $r_s=3$ is accepted). However, in general, this case is unreal. Second, according to the physical meaning of exact Eq.(16) [30], the core sample volume under drying, shrinkage, and measurement is assumed to only include the soil matrix; that is, cracks inside the sample are assumed not to develop. However, the crack development in cores strongly depends on measurement conditions (sample size, drying regime, and others), and is usually uncontrolled. For this reason, ignoring the possible crack volume inside the core at finding the $r_s(W)$ dependence (and the corresponding total crack volume dependence), even from continuous measuring of the volume and subsidence of cylindrical soil samples at the water content decrease, can, in general, give $r_s(W)$ with essential inaccuracy. The third and final assumption is that the subsidence of a core sample is equal to the subsidence of a real cracked, but connected soil layer in field conditions [30, 31]. This case is obviously unreal and leads to the essential inaccuracy of the $r_s(W)$ dependence for a soil layer (and the corresponding total crack volume dependence).

The above analysis [6, 7] also showed that for obtaining the correct $r_s$ values and correct crack volume estimates one also needs to know the shrinkage curve, $\overline{V}(W)$ of the soil matrix without cracks, in addition to measurements with cores. The recent model of the reference shrinkage curve [8, 9] permits one to predict the soil matrix shrinkage curve, $\overline{V}(W)$ for an aggregated soil with any clay content. Then, using results from [6, 7] one can estimate two variants of the corrected $r_s(W)$ dependence through three shrinkage curves: the curve $\overline{V}_1'(W)$ of a soil layer with cracks in Bronswijk's approximation (the initial layer is composed of contacting, but disconnected cubes); the curve $\overline{V}_1(W)$ of a real connected soil layer with cracks; and the curve $\overline{V}(W)$ of a soil matrix without cracks. Given $\overline{V}_1'(W)$, $\overline{V}_1(W)$, and $\overline{V}(W)$ (as well as $\overline{V}_o = \overline{V}(W_o)$, $W_o$ is the maximum swelling point before shrinkage starts), the corrected shrinkage geometry factor of a soil core, $r_{sM}$ is calculated as

$$r_{sM}(W) = \log(\overline{V}(W)/\overline{V}_o)/\log(\overline{V}_1'(W)/\overline{V}_o) \ ; \tag{21}$$

and the corrected shrinkage geometry factor of a soil layer, $r_s$ is calculated as

$$r_s(W) = \log(\overline{V}(W)/\overline{V}_o)/\log(\overline{V}_1(W)/\overline{V}_o) \ . \tag{22}$$

For the case of a soil layer (i.e., with $r_s(W)$ from Eq.(22)) Eq.(17) and (19) are modified as

$$\delta_T(W) = 1 - [1 - \delta_V(W)]^{1/r_s(W)} \tag{23}$$

and

$$\delta_A(W) = 1 - [1 - \delta_V(W)]^{[1-1/r_s(W)]} \ , \tag{24}$$

respectively, where

$$\delta_V(W) = 1 - \overline{V}(W)/\overline{V}_o \ . \tag{25}$$

In this case Eq.(20) is replaced with

$$\delta_A(W) = 1 - [1 - \delta_T(W)]^{[r_s(W)-1]} \ . \tag{26}$$

For the case of a soil core one should use in Eq.(23), (24), and (26) $r_{sM}(W)$ from Eq.(21) instead of $r_s(W)$ from Eq.(22). The calculation of $\overline{V}_1(W)$ (i.e., for the case of soil layer) from the experimental data is illustrated in Materials and Methods.

Thus, the modification flowing out of the works [6-9] is reduced to the calculation of the vertical and horizontal crack characteristics that are connected with shrinkage (Eq.(3)-(15); see also [1-3]) using: (i) the reference shrinkage curve, $\overline{V}(W)$ as soil matrix shrinkage curve in Eq.(25) instead of usual shrinkage curve; (ii) the $r_s$ factor calculated from Eq.(21) or (22) as a function of $W$ instead of $r_s=3$; and (iii) modified relations between $\delta_T(W)$, $\delta_A(W)$, and $\delta_V(W)$ from Eq.(23), (24), and (26) instead of Eq.(17), (19), and (20), respectively.

*Specification of the Expression for Horizontal Surface Shrinkage*

The exact expression for $\delta_A(z)$ is

$$\delta_A(z) = R(z,z_m)L(z_m) + \int_{z_m}^{z} R(z,h) dL(h) \ . \tag{27}$$

This expression is one from Eq.(8) with an additional term, $R(z,z_m)L(z_m)$ which gives the contribution of the largest cracks of depth $z_m$. Their opening at depth $z$ is $R(z,z_m)$. According to Eq.(3) and (6) the specific length, $L(z_m)$ of the traces of such cracks at depth $z_m$ is

$$L(z \to z_m) \cong 0.45/z_o \, , \tag{28}$$

At usual $z_o \cong 40$-$60$ cm and $R(z,z_m) \leq 1$ cm (see [1]) the $R(z,z_m)L(z_m)$ term, that is, the contribution of the largest cracks in Eq.(27) is negligible. However, at sufficiently small $z_o$ (see Materials and Methods) the $R(z,z_m)L(z_m)$ term can be essential.

*Specification Connected with the Point of Reference on the Soil Depth Axis*

Usually soil subsidence is determined at any moment using the depth axis $z_I$ (Fig.**1**) with $z_I$=0 at the *initial* position of the soil surface. Cracks obviously are inside soil at any shrinkage. For this reason different crack characteristics (unlike subsidence characteristics) are naturally determined at any time moment using the depth axis $z_C$ (Fig.**1**) with $z_C$=0 at the *current* position of the soil surface. Correspondingly, by default, we imply that all the depth coordinates ($z_m$, $z$, $h$, $z'$, and $h'$) in Eq.(7)-(10), and (15) relating to the crack characteristics are of type $z_C$ and in Eq.(11)-(14) relating to the subsidence characteristics are of type $z_I$. However, both the subsidence and crack characteristics can be expressed through either $z_C$ or $z_I$ depth because $z_C$=$z_C(z_I)$. This dependence is determined by the following equation and initial condition (see Fig.**1**) as

$$dz_C/dz_I = 1 - \delta_T(z_I) \, , \qquad z_C\big|_{z_I = 0} = 0 \tag{29}$$

that immediately flow out of the definition and physical meaning of $z_C$, $z_I$, and $\delta_T$ (see above).

*Relations between Different Forms of Shrinkage*

Equation (29) and the volume balance at shrinkage with vertical crack development allows one to derive a useful relation between $\delta_V$, $\delta_A$, and $\delta_T$. The volume decrease $\Delta V_m$ of the soil matrix layer between depths $z_{mI}$ and $z_I$ (per unit area), the subsidence $S$ of the soil at depth $z_I$, and the cumulative volume of vertical cracks, $V_{v\,cr}$ upwards from depth $z_{mC}$ to $z_C$ (per unit area), are connected by the volume balance relation as

$$\Delta V_m(z_I) = S(z_I) + V_{v\,cr}(z_C) \, . \tag{30}$$

By definition of $\Delta V_m$, $S$, and $V_{v\,cr}$ one can write them as

$$\Delta V_m(z_I) = \int_{z_I}^{z_{mI}} \delta_V(z) dz \, ; \tag{31}$$

$S$ according to Eq.(11) with $z \to z_I$ and $z_m \to z_{mI}$, and (accounting for Eq.(29))

$$V_{v\,cr}(z_C) = \int_{z_C}^{z_{mC}} \delta_A(z) dz = \int_{z_I}^{z_{mI}} \delta_A(z)(1 - \delta_T(z)) dz = V_{v\,cr}(z_I) \, . \tag{32}$$

Using Eq.(31), (11), and (32) the integral relation from Eq.(30) can be rewritten in differential form as

$$\delta_V(z_I) = \delta_T(z_I) + \delta_A(z_I)(1 - \delta_T(z_I)) \, . \tag{33}$$

Here $z_I$ can be changed by water content $W(z_I)$ (see the text before Eq.(16)). One can check that replacement $\delta_T$ and $\delta_A$ in Eq.(33) from Eq.(23) and (24) leads to an identity. This is natural because Eq.(23) and (24) were also obtained from the volume balance relation in the form of Eq.(16). Obtaining Eq.(33) in two different ways confirms Eq.(11) and (32) for $S(z_I)$ and $V_{v\,cr}(z_I)$, respectively, that will be needed below.

*An Additional Volume Shrinkage of the Soil Matrix due to Horizontal Cracks*

According to [2, 3] horizontal cracks develop after vertical ones and without additional subsidence and change of the vertical-crack volume. For this reason the additional volume decrease $\Delta'V_m$ of the soil matrix layer between depths $z_{mI}$ and $z_I$ (per unit area) is equal to the cumulative volume of the horizontal cracks, $V_{h\,cr}$ upwards from depth $z_{mC}$ to $z_C$ (per unit area) as

$$\Delta' V_m(z_I) = V_{h\,cr}(z_C) \ . \tag{34}$$

Similar to Eq.(31) one can write $\Delta' V_m(z_I)$ as

$$\Delta' V_m(z_I) = \int_{z_I}^{z_{mI}} \delta'_V(z) \mathrm{d}z \tag{35}$$

where $\delta'_V$ is an additional specific volume shrinkage of the soil matrix that leads to development of horizontal cracks. Note that $\delta_V$ from Eq.(31) corresponds to a homogeneous drying and shrinkage over the soil volume. Unlike that $\delta'_V$ from Eq.(35) corresponds to averaging the inhomogeneous shrinkage (namely, an additional shrinkage of the thin drying soil layers along the walls of vertical cracks) over the total soil volume. Similar to Eq.(32) and using Eq.(15) one can write the cumulative volume of the horizontal cracks, $V_{h\,cr}(z_C)$ as

$$V_{h\,cr}(z_C) = \int_{z_C}^{z_{mC}} v_{h\,cr}(z)\mathrm{d}z = \int_{z_I}^{z_{mI}} v_{h\,cr}(z)(1-\delta_T(z))\mathrm{d}z = -\int_{z_I}^{z_{mI}} \frac{\mathrm{d}\overline{\Delta S}(z)}{\mathrm{d}z}(1-\delta_A(z))\mathrm{d}z = V_{h\,cr}(z_I) \ . \tag{36}$$

Accounting for Eq.(35) and (36) the integral relation from Eq.(34) can be rewritten in differential form as

$$\delta'_V(z_I) = v_{h\,cr}(z_I)(1-\delta_T(z_I)) = -\frac{\mathrm{d}\overline{\Delta S}(z_I)}{\mathrm{d}z_I}(1-\delta_A(z_I)) \ . \tag{37}$$

Finally, the sum of Eq.(30) and (34) gives the total volume balance accounting for the total volume decrease $DV_m$ of the soil matrix layer between depths $z_{mI}$ and $z_I$ (per unit area), soil surface subsidence, $S$ of the soil at depth $z_I$, and cumulative volume of both the vertical ($V_{v\,cr}$) and horizontal ($V_{h\,cr}$) cracks upwards from depth $z_{mC}$ to $z_C$ (per unit area) as

$$DV_m(z_I) = \Delta V_m(z_I) + \Delta' V_m(z_I) = S(z_I) + V_{v\,cr}(z_C) + V_{h\,cr}(z_C) \ . \tag{38}$$

Note that only the partial volume balance with vertical cracks (Eq.(30)) corresponds to Bronswijk's approximation [30] (Eq.(16)).

## MATERIALS AND METHODS

**Data Used**

Data that would permit one to totally illustrate the possibilities of the modified approach to the geometry of soil crack networks are not available. The most suitable are data of the lysimeter experiment from [40]. These data have been considered from the viewpoint of a crack network volume [1]. We intend analyze these data anew, but in more detail and accounting for the above indicated modifications. Bronswijk [40] investigated a heavy clay soil from the central part of the Netherlands. The height of the undisturbed large soil core was 60 cm and the diameter 27.4 cm. Initially the large core was water saturated. Then, for 82 days the ground water level was kept constant at 55 cm below the initial position of the soil surface and water evaporated only from the bare soil surface. We used the following experimental data from this experiment: (i) the shrinkage curve of aggregates; (ii) gravimetric water content vs. drying duration of the upper soil layer of 7.5 cm thickness; (iii) water content - soil depth profiles at drying durations of 33, 39, and 82 days; (iv) the subsidence of the soil surface in the lysimeter vs. drying duration, in particular after 33, 39, and 82 days; and (v) the total crack volume (of vertical and horizontal ones) measured with a direct method after drying for 82 days. In the data analysis we also used values of a number of separate parameters from this experiment (Table **1**). They will be discussed below.

The following preliminary remarks are necessary. All crack and subsidence characteristics, that we are interested in, depend on soil water content, $W$. The latter depends on drying duration, $t$ and soil depth, $z$. We use

*W*(*t*) dependence at *z*=3 cm from Fig.4B of [40] to connect (numerically) the water content, *W* and drying duration, *t*, and *W*(*z*) dependence at *t*=33, 39, and 52 days from Fig.5 of [40] to connect *W* and soil depth, *z*.

**Data Analysis**

Using the modified model one can, in detail, estimate different distributions relating to vertical and horizontal crack networks (based on Eq.(9) and (10)). For comparison, however, there are only data on the subsidence of the soil surface after drying for 33, 39, and 82 days as well as cumulative (vertical and horizontal) crack volume per unit area after drying for 82 days [40]. For this reason we only estimated the model predicted values for the subsidence of the soil surface as well as the total crack volume and contributions of the vertical and horizontal cracks for the above three drying durations.

To find the vertical and horizontal crack network characteristics as well as the soil surface subsidence of the large core one should first estimate the specific volume shrinkage of the soil matrix (without cracks), $\delta_V(W)$ (Eq.(25)), horizontal surface shrinkage, $\delta_A(W)$ (Eq.(24)), and vertical linear shrinkage, $\delta_T(W)$ (Eq.(23)) for the soil. As stated above, $\delta_V(W)$ is estimated by the reference shrinkage curve, $\overline{V}(W)$ (Eq.(25)). This reference shrinkage curve was not measured by Bronswijk [40]. The algorithm of its estimation is considered in the following subsection. To find $\delta_A(W)$ and $\delta_T(W)$ (with the known $\delta_V(W)$ and accounting for above modifications) one needs the shrinkage geometry factor, $r_s(W)$ for a soil layer (Eq.(24) and (23)). In turn, to estimate $r_s(W)$ (Eq.(22)) one needs (in addition to the reference shrinkage curve of the soil, $\overline{V}(W)$) the relative shrinkage curve $\overline{V}_1(W)/\overline{V}_o$ of the soil layer with cracks. The estimation algorithm of $\overline{V}_1(W)/\overline{V}_o$, $r_s(W)$, $\delta_A(W)$, and $\delta_T(W)$ is considered below in the corresponding subsection. Then numerical estimation of $\overline{V}(W)$ and $\overline{V}_1(W)/\overline{V}_o$ based on available data, and of $\delta_V(W)$, $\delta_A(W)$ and $\delta_T(W)$ are considered.

*Estimation of the Soil Reference Shrinkage Curve* ($\overline{V}(W)$)

The soil reference shrinkage curve describes shrinkage of a soil matrix without cracks [8-10]. Sometimes, at high clay content, the shrinkage curve of a clayey paste, or separate soil aggregates, or small clods can be used as the reference shrinkage curve (see examples in [7]). In general, however, the reference shrinkage curve does not reduce to the shrinkage curve of a clayey paste or soil aggregates even at high clay content. The clay content in the lysimeter experiment [40] was high (Table **1**). In this case the reference shrinkage curve is determined by six physical soil parameters [8]: oven-dried specific volume ($\overline{V}_z$), maximum swelling water content ($W_o$), mean solid density ($\rho_s$), soil clay content (*c*); oven-dried structural porosity ($P_z$), and the ratio of an aggregate solid mass to the solid mass of an intraaggregate matrix (*K*>1).

All of the above parameters can be found using data from [40]. We took the *c* value (Table **1**) as a mean for the Bruchen heavy clay from Bronswijk and Evers-Vermeer [41]. We took $P_z$=0 (Table **1**) since, from Fig.6 of [40], the crack volume including the interaggregate (structural) pores is negligible at maximum swelling, The maximum gravimetric water content of the soil, $W_o$ (Table **1**) follows from Fig.4B of [40]. To estimate the $\rho_s$ value the shrinkage curve of soil aggregates from Fig.2A of [40] was preliminarily recalculated from moisture ratio ($\theta$) - void ratio (*e*) coordinates to gravimetric water content (*w*) - specific volume ($\overline{V}_a$) coordinates (Fig.**2** and **3**), using relations $w=(\rho_w/\rho_s)\theta$ and $\overline{V}_a=(e+1)/\rho_s$ where $\rho_w$ is the water density. Then, $\rho_s$ for the soil (Table **1**) follows from the correspondence between the maximum moisture ratio of aggregates, $\theta \cong 1.15$ cm$^3$ cm$^{-3}$ and their maximum gravimetric water content, $w_o$. The latter is equal to the maximum gravimetric water content of the soil, $w_o=W_o$ (Table **1**; Fig.**2** and **3**) because according to Fig.6 of [40], the crack volume including interaggregate pores is negligible at maximum swelling.

Accounting for the high soil clay content the aggregate shrinkage curve, $\overline{V}_a(w)$ (indicated above; Fig.**2** and **3**) can be considered as the shrinkage curve of the intraaggregate matrix of the soil (cf., [8]). Then the oven-dried specific volume of the intraaggregate matrix, $\overline{V}_{az}$ (Table **1**; Fig.**2** and **3**) follows from Fig.2A of [40] (after transformation to *w* and $\overline{V}_a$ coordinates).

According to [8] $\overline{V}_z$ (Fig.**2** and **3**) can be expressed through $\overline{V}_{az}$ (Fig.**2** and **3**; Table **1**) and *K* (Table 1; see below) as

$$\overline{V}_z = \overline{V}_o \ (K-1)/K + \overline{V}_{az}/K \tag{39}$$

where $\bar{V}_o = \bar{V}_{ao}$ (Fig.**2** and **3**) is expressed through other known values [8] and, in addition, immediately follows from Fig.2A of [40].

Finally, note that there are two types of the reference shrinkage curve, $\bar{V}(W)$ (Fig.**2** and **3**) depending on its behavior in the structural shrinkage area [8, 9]. Each of these possibilities of $\bar{V}(W)$ is characterized by its $K$ value (estimating the $K$ ratio is discussed in one of the following subsections). After estimation of the $\bar{V}(W)$ curve the specific volume shrinkage, $\delta_V(W)$ is found from Eq.(25).

***Estimation of the Relative Shrinkage Curve of a Soil Layer with Cracks*** ($\bar{V}_l(W)/\bar{V}_o$), ***the Shrinkage Geometry Factor*** ($r_s(W)$), ***and Specific Shrinkage*** ($\delta_A(W)$ and $\delta_T(W)$)

The following preliminary remarks are necessary. The $\bar{V}_l(W)/\bar{V}_o$ ratio is needed to estimate $r_s(W)$. However, estimating $r_s(W)$ at a given water content of a soil layer, we cannot use the shrinkage of the large core in the lysimeter at a given drying duration because of the essential change of the water content, and correspondingly $r_s$, with depth in the large core. Note that Bronswijk [40] could use Eq.(16) for the lysimeter core as a whole since he accepted $r_s$=const=3. In general, the initial layer thickness, $T_o$ in Eq.(16) should be such that the $r_s(W)$ variation inside the layer (after a given drying duration) was relatively small compared to a mean $r_s$ value in the layer. Otherwise, the use of Eq.(16) is not correct with $r_s$ depending on $W$. For this reason the layer should be sufficiently thin compared with the large core. We consider the upper soil layer of 7.5 cm thickness as such a layer with sufficiently homogeneous water content over the layer volume at any given time moment,. In addition, we use this relatively thin layer because the total range of the variation of the mean water content in the layer during drying (for 82 days) essentially exceeds that in the deeper layers (see Fig.4B of [40]). Note also that the small layer thickness and coupling with lower layers of the large core allows one to consider its shrinkage as that of a layer, but not of a small core, and correspondingly to use Eq.(22), but not Eq.(21).

Thus, the $\bar{V}_l(W)/\bar{V}_o$ ratio that is needed to estimate the shrinkage geometry factor for the soil layer, $r_s(W)$ (Eq.(22)) is

$$\bar{V}_l(W)/\bar{V}_o = T(W)/T_o \qquad (40)$$

where $T(W)$ `and $T_o$ (Table **1**) are as defined above, the current and initial thickness of the upper lysimeter layer, respectively. One can write $T(W)$ as

$$T(W)=T_o-nS(W), \qquad 0<W<W_o \qquad (41)$$

where $S(W)$ is the experimental soil surface subsidence from Fig.6 of [40] (that reaches 12.4 mm through 82 days) as a function of the mean water content, $W$ of the upper lysimeter layer from Fig.4B of [40]. We assume that $n$ does not depend on $W$ (or drying duration). This will be checked. In Eq.(41) $n<1$ because of some subsidence of the bottom surface of the upper lysimeter layer during drying. At the same time $n>0$ because the bottom surface subsidence in any case is less than $S(W)$.

After estimating $\bar{V}(W)$ (previous subsection) and $\bar{V}_l(W)/\bar{V}_o$ (at given $n$ and $K$) $r_s(W)$, $\delta_A(W)$, and $\delta_T(W)$ are found from Eq.(22)-(24). Estimating the $n$ and $K$ values is discussed in the following subsection.

***Numerical Estimation of Introduced Parameters*** (*n and K*) *from Available Data*

After estimation of $\delta_V(W)$, $\delta_A(W)$, and $\delta_T(W)$ (at given $n$ and $K$) one finds the soil surface subsidence and the crack network characteristics of the large core from Eq.(3)-(15) and modifications that are summarized (for the case of the lysimeter experiment) in Eq.(32) and (36). Together with that, we used the experimental dependencies from Fig. 4B, 5, and 6 of [40] to connect $t$, $z$, and $W$ values.

Using the above algorithms we estimated the $n$ coefficient, equalizing the predicted soil surface subsidence in the lysimeter after drying for 82 days [see Eq.(11) at $z$=0 and $\delta_T(z)=\delta_T(W(z, t$=82d$))$] and the experimental subsidence after $t$=82d from Fig.6 of [40] (see Table **2**). Then the $n$ value found (Table **1**) was used to find subsidence for $t$=33 and 39d (Table **2**).

In general, the $K$ value can be estimated from data on soil structure and texture [42]. However such data are lacking in the case under consideration. The $K$ ratio was estimated (at the found $n$), equalizing the predicted total cumulative crack volume (of vertical and horizontal those) per unit area of the lysimeter after drying for 82 days [see Eq.(32) and (36) at $z_l$=0, $\delta_T(z)=\delta_T(W(z, t=82d))$, and $\delta_A(z)=\delta_A(W(z, t=82d))$], and the experimental total cumulative crack volume from Fig.6 of [40] (see Table **3**). The $n$ and $K$ values found (Table **1**) were used to find the crack volume for 33 and 39d (Table **3**).

## RESULTS AND DISCUSSION

Table **1** shows the $n$ value that was estimated, as stated above, at 82 d drying duration. Table **2** shows the experimental subsidence of the soil surface in the lysimeter at 33, 39, and 82 d drying duration from Bronswijk [40] and the model prediction for these drying durations with the indicated $n$ value. It is worth noting, first, that the discrepancies between experimental and predicted soil surface subsidence values are within the limits of experimental errors (<~1 mm) for any drying duration (not only for 82 days) at the same $n$ value. This means that the presentation of Eq.(41) is reasonable. Second, the $n$ value that is close to unity (Table **1**) shows that the subsidence of the lower surface of the upper 7.5 cm layer is small compared with that of the soil surface (i.e., $(1-n)S(W) \ll S(W)$). This is in agreement with very weak water content variation at larger lysimeter depths [40]. Note, in addition, that $n$=1 leads to the predicted soil surface subsidence values (for different drying durations) that are not in agreement with observations. This means that the small deflection of $n$ from unity is of principle importance.

Table **1** shows the $K$ values that were estimated, as stated above, at 82 d drying duration for two of the possible variants of the reference shrinkage curve of the soil, $\overline{V}(W)$ (Fig.**2** and **3**). First, a small deflection of the estimated $K$ values of unity for both types of the reference shrinkage curves (Fig.**2** and **3**) should be noted. This corresponds to the modeling of the reference shrinkage curve at sufficiently high soil clay content [8] and is in the agreement with the estimated $K$ values for a number of other real soils with high clay content [8]. Note, however, that $K$=1 (when the reference shrinkage curve, $\overline{V}(W)$ in Fig.**2** and **3** are replaced with $\overline{V}_a(W)$) leads to unsatisfactory results. The corresponding total crack volume (per unit area), $V_{cr}=V_{v\,cr}+V_{h\,cr}$~34 mm turns out to be rather more than the observed value $V_{cr}$~23 mm (Table **3**). This means that the small deflection of $K$ from unity is of principle importance. Second, in spite of the proximity of both the estimated $K$ values (Table **1**) to unity, their small variations with $\Delta K=\pm 0.025$ leads to the essential change of the corresponding estimated total crack volume for 82 d drying duration compared to the observed value, $V_{cr}$~23 mm (Table **3**). This change is out of the limits of experimental errors (>1 mm) and demonstrates the appreciable effect of the intraaggregate soil structure (which is expressed by the $K$ value) on the shrinkage crack characteristics.

Figure **4** shows the model predicted shrinkage geometry factor of the upper soil layer with the initial thickness of 7.5 cm as a function of the (mean) water content in the layer from the lysimeter experiment [40] for the two possible types of reference shrinkage curve (see Fig.**2** and **3**). Unlike $r_s$=const=3, that was postulated in [40], $r_s$ demonstrates an initial sharp increase up to $r_s$~35-45 with decrease in water content. This increase is stipulated by the absence of soil surface subsidence for approximately the first two-three days of drying (see Fig.6 of [40]). In such conditions the volume shrinkage of the soil matrix in the layer can only be realized through a quick crack formation that corresponds to the sharp increase of $r_s$ with initial drying. Then with the appearance and increase of soil surface subsidence, $r_s$ quickly decreases to approximately ten for the next two-three days (at $W$~0.35 kg kg$^{-1}$ in Fig.**4**), and in ten days of drying decreases to ~1.5 (at $W$~0.3 kg kg$^{-1}$ in Fig.**4**). With the subsequent drying the $r_s$ decrease is relatively small ($W$~0.25 kg kg$^{-1}$ in Fig.**4** corresponds to drying for ~36 days). Qualitatively, such $r_s(W)$ behavior with a more or less sharp maximum and following approximate constancy at drying is similar to that from Fig.8 of [7] for another soil of high clay content.

Figure **5** illustrates the predicted depth dependence of the cumulative (upward from depth $z_m$) vertical and horizontal crack volume as well as the total crack volume (per unit area) after 82d drying duration for the case of the reference shrinkage curve in Fig.3. In addition, the predicted $V_{v\,cr}$, $V_{h\,cr}$, and $V_{cr}=V_{v\,cr}+V_{h\,cr}$ volumes (per unit area) at the soil surface, $z_C$=0 for three drying durations are given in Table **3** for the two possible types of reference shrinkage curve (from Fig.**2** and **3**). For comparison Table **3** also shows available data for the total crack volume from [40] as well as the prediction from [40] and [1]. First, it should be noted that the relatively small (for this soil) estimated contribution of the horizontal cracks to the total crack volume (at $z_C$=0 and at different drying duration in Table **3**; or at different depths and after drying for 82d in Fig.**5**), is, nevertheless, interesting because of the possible contribution of the cracks to the soil hydraulic conductivity. Second, for the soil under consideration the differences between $V_{v\,cr}$, $V_{h\,cr}$, and $V_{cr}$ (Table **3**) predicted for the two possible variants of the reference shrinkage curve (Fig.**2** and **3**) are within the limits of the experimental errors (<~1 mm) for any drying duration. Third, Table **3** shows that the estimates of this work for the vertical crack volume, $V_{v\,cr}$ are essentially more accurate compared to similar estimates in [1]. The difference between the estimates of $V_{v\,cr}$ in this work and [40] is also appreciable at 33 and 39 d drying duration. The reason for this is the use in [40] and [1] of the postulated value, $r_s$=const=3. Unlike that, in this work the corrected $r_s(W)$ dependence was considered and used.

Finally, it is worth reiterating that the modified approach permits one to obtain, in addition to the cumulative crack volume, not only the specific crack volume and separately the vertical and horizontal crack volume, but also the different relevant distributions of crack characteristics. They were not estimated here because of the lack of corresponding experimental data.

## CONCLUSION

Predicting soil crack network geometry is important for different applications of soil hydrology. There are a number of approaches that deal with the soil crack network. The approach from [1-4] seems to be the most physically substantiated and to give the most detailed information about the soil crack network characteristics. The main drawback of this approach is connected with the use of a non-accurate shrinkage geometry factor from Bronswijk [30, 31] in the estimation of the soil crack volume. Recent works [6, 7] showed that some implicit assumptions of [30, 31] are violated in real conditions, and suggested ways to calculate the correct shrinkage geometry factor. In such calculations one also needs the shrinkage curve of the soil matrix without cracks. For such a shrinkage curve one can use the so-called reference shrinkage curve [8-10]. In this work we modify the approach to the prediction of soil crack network geometry from [1-4], accounting for the results from [6-10] and a number of specifications. Then, we validate and illustrate the modified approach analyzing available data from Bronswijk [40]. The obtained results permit one to estimate different crack network characteristics of shrinking soils with essentially higher accuracy.

## NOTATION

| | |
|---|---|
| $A_o$ | initial total area of a horizontal uncracked soil cross-section at a depth $z$, m$^2$ |
| $A(z)$ | current total area of a horizontal uncracked soil cross-section at a depth $z$, m$^2$ |
| $C$ | crack connectedness, dimensionless |
| $c$ | clay content, dimensionless |
| $DV_m$ | total volume decrease of the soil matrix layer between depths $z_m$ and $z$ (per unit area), m$^3$ m$^{-2}$ |
| $d$ | average crack spacing, m |
| $e$ | void ratio, m$^3$ m$^{-3}$ |
| $f(x)$ | probability of connection of cracks of any orientation of dimension $<x$ (or volume fraction of fragments of all the dimensions $<x$), dimensionless |
| $f_m = f(x_m)$ | volume fraction occupied by all the fragments, or the fragment formation probability, dimensionless |
| $h$ | crack tip depth, m |
| $h'$ | integration variable, m |
| $K$ | ratio of an aggregate solid mass to the solid mass of an intraaggregate matrix, dimensionless |
| $L$ | total length of vertical-crack traces on a horizontal cross-section at depth $z$, m$^{-1}$ |
| $n$ | coefficient in Eq.(41), dimensionless |
| $P(z, h)$ | cumulative fraction of the total specific crack volume, or of the total specific crack cross-sectional area at a depth $z$, related to cracks with tips at depths $>h$, dimensionless |
| $P_z$ | oven-dried structural porosity, dimensionless |
| $R(z, h)$ | mean width at depth $z$ of cracks with tips at depth $h$, m |
| $r_s$ | shrinkage geometry factor of a soil layer, dimensionless |
| $r_{sM}$ | corrected shrinkage geometry factor of a soil core, dimensionless |
| $S(z)$ | subsidence of the drying soil along a vertical elevation not containing a vertical crack as a function of soil depth, $z$; $S(z) \equiv S(W(z))$, m |
| $s$ | measurable mean crack spacing at the soil surface, m |
| $T(z)$ | current thickness of a horizontal soil layer around a depth $z$, m |
| $T_o$ | initial thickness of a horizontal soil layer around a depth $z$ and, in particular, of the upper 7.5 cm layer of the lysimeter, m |
| $t$ | drying duration, day |
| $V(z)$ | current volume of a soil matrix (without cracks) in a horizontal soil layer around a depth $z$, m$^3$ |
| $V_o$ | initial volume of a soil matrix (without cracks) in a horizontal soil layer around a depth $z$, m$^3$ |
| $V_{cr}$ | cumulative crack volume in the lysimeter (per unit area), m$^3$ m$^{-2}$ |
| $V_{h\,cr}$ | cumulative horizontal crack volume in the lysimeter (per unit area), m$^3$ m$^{-2}$ |
| $V_{v\,cr}$ | cumulative vertical crack volume in the lysimeter (per unit area), m$^3$ m$^{-2}$ |
| $\overline{V}(W)$ | shrinkage curve of the soil matrix (without cracks) or the soil reference shrinkage curve, dm$^3$ kg$^{-1}$ |
| $\overline{V}_a(w)$ | shrinkage curve of an intraaggregate matrix, dm$^3$ kg$^{-1}$ |

$\overline{V}_{az}$     oven-dried specific volume of an intraaggregate matrix ($\overline{V}_a(w_z)$) per unit mass of the matrix itself, dm$^3$ kg$^{-1}$

$\overline{V}_{ao}$     $\overline{V}_a(w_o)$ value, dm$^3$ kg$^{-1}$

$\overline{V}_o$     $\overline{V}(W_o)$ value, dm$^3$ kg$^{-1}$

$\overline{V}_1(W)$     shrinkage curve of a real connected soil layer with cracks, dm$^3$ kg$^{-1}$

$\overline{V}_z$     $\overline{V}(W_z)$ value, dm$^3$ kg$^{-1}$

$\overline{V}_1'(W)$     shrinkage curve of a soil layer with cracks in Bronswijk's approximation, dm$^3$ kg$^{-1}$

$v_{h\,cr}(z)$     specific width of the horizontal cracks per unit height of a vertical profile or the specific volume of the horizontal cracks (per unit volume of soil), dimensionless

$W(z)$     current soil water content profile (i.e., for a given drying duration), kg kg$^{-1}$

$W$     total gravimetric water content (per unit mass of oven-dried soil), kg kg$^{-1}$

$W_n$     total gravimetric water content at the endpoint of the basic shrinkage area of a soil, kg kg$^{-1}$

$W_s$     total gravimetric water content at the endpoint of the structural shrinkage area of a soil, kg kg$^{-1}$

$W_z$     total gravimetric water content at the shrinkage limit of a soil, kg kg$^{-1}$

$W_o$     maximum swelling point of a soil before the start of shrinkage, kg kg$^{-1}$

$w$     water content of an intraaggregate matrix (per unit mass of oven-dried intraaggregate matrix itself), kg kg$^{-1}$

$w_n$     gravimetric water content of an intraaggregate matrix at the endpoint of the basic shrinkage of the intraaggregate matrix, kg kg$^{-1}$

$w_z$     gravimetric water content of an intraaggregate matrix at the shrinkage limit of the intraaggregate matrix, kg kg$^{-1}$

$w_o$     gravimetric water content of an intraaggregate matrix at the maximum swelling point of the intraaggregate matrix; $w_o = W_o$, kg kg$^{-1}$

$x_m = 4d$     maximum dimension of fragments, m

$z$     soil depth, m

$z_C$     coordinate of soil depth at shrinkage with movable point of reference at a *current* position of the soil surface, m

$z_I$     coordinate of soil depth at shrinkage with fixed point of reference at the *initial* position of the soil surface, m

$z_o$     thickness of an upper soil layer (a few tens of centimeters) of intensive cracking, m

$z_m$     maximum crack depth, m

$z_1, z_2$     depths of the upper and lower boundaries of a soil layer, m

$z'$     integration variable, m

$\Delta A(z) \equiv A_o - A(z)$     increment of the uncracked area under shrinkage at a given depth $z$, m$^2$

$\Delta S(z, h)$     potential relative subsidence at depth $z$ of a vertical profile containing a vertical crack of depth $h$, m

$\overline{\Delta S(z)}$     mean potential relative subsidence (MPRS), m

$\Delta T(z) \equiv T_o - T(z)$     increment of the layer thickness under shrinkage and cracking around a given depth $z$, m

$\Delta V(z) \equiv V_o - V(z)$     increment of the soil matrix volume under shrinkage and cracking around a given depth $z$, m$^3$

$\Delta V_m$     volume decrease of the soil matrix layer between depths $z_m$ and $z$ (per unit area) due to soil subsidence and cumulative volume of vertical cracks, m$^3$

$\Delta' V_m$     additional volume decrease of the soil matrix layer between depths $z_m$ and $z$ (per unit area) due to cumulative volume of horizontal cracks, m$^3$

$\delta_A(z) \equiv \Delta A(z)/A_o$     specific horizontal surface shrinkage per initial unit area (or the total specific crack cross-section area at depth $z$), dimensionless

$\delta_{cr}(z, h)$     linear vertical shrinkage at a point on the wall of a vertical crack as a function of the crack tip depth, $h$, and the depth of the point on the wall, $z \leq h$, dimensionless

$\delta_T(z) \equiv \Delta T(z)/T_o$     specific linear vertical shrinkage of the soil (per initial unit layer thickness) at depth $z$, dimensionless

$\delta_V(z) \equiv \Delta V(z)/V_o$     specific volume shrinkage of a soil matrix, dimensionless

$\delta'_V$     additional specific volume shrinkage of a soil matrix that leads to the development of horizontal cracks, dimensionless

$\theta$     moisture ratio, m$^3$ m$^{-3}$

$\pi(z, h)$     differential fraction (probability density) of the total specific crack volume (per unit volume of soil), or of the total specific crack cross-sectional area (per unit cross-sectional area), at a depth $z$ related to cracks with tips in a unit interval at depth $h$, dimensionless

$\rho_w$     water density, g cm$^{-3}$
$\rho_s$     mean density of soil solids, g cm$^{-3}$
$\omega$     degree index in Eq.[3], dimensionless

## REFERENCES

[1] Chertkov VY, Ravina I. Modeling the crack network of swelling clay soils. Soil Sci Soc Am J 1998; 62: 1162-1171.
[2] Chertkov VY, Ravina I. Morphology of horizontal cracks in swelling soils. Theoretical and Applied Fracture Mechanics 1999; 31: 19-29.
[3] Chertkov VY, Ravina I. Analysis of the geometrical characteristics of vertical and horizontal shrinkage cracks. J Agric Engng Res 1999; 74: 13-19.
[4] Chertkov VY. Using surface crack spacing to predict crack network geometry in swelling soils. Soil Sci Soc Am J 2000; 64: 1918-1921.
[5] Chertkov VY. Modelling cracking stages of saturated soils as they dry and shrink. Europ J Soil Sci 2002; 53: 105-118.
[6] Chertkov VY, Ravina I, Zadoenko V. An approach for estimating the shrinkage geometry factor at a moisture content. Soil Sci Soc Am J 2004; 68: 1807-1817.
[7] Chertkov VY. The shrinkage geometry factor of a soil layer. Soil Sci Soc Am J 2005; 69: 1671-1683.
[8] Chertkov VY. The reference shrinkage curve at higher than critical soil clay content. Soil Sci Soc Am J 2007; 71(3): 641-655.
[9] Chertkov VY. The soil reference shrinkage curve. Open Hydrology Journal 2007; 1: 1-18.
[10] Chertkov VY. The reference shrinkage curve of clay soil. Theor. and Appl. Fracture Mechanics 2007; 48(1): 50-67.
[11] Gdoutos EE. Fracture Mechanics. London. Kluwer Academic Publishers. 1993.
[12] Hallett PD, Newson TA. Describing soil crack formation using elastic-plastic fracture mechanics. Europ J Soil Sci 2005; 56: 31-38.
[13] Bazant PZ, Cedolin L. Stability of Structures, Oxford University Press. 1991.
[14] Jenkins DR. Optimal spacing and penetration of cracks in a shrinking slab. Physical Review E 2005; 71: 056117 (1-8).
[15] Yoshida S, Adachi K. Numerical analysis of crack generation in saturated deformable soil under row-planted vegetation. Geoderma 2004; 120: 63-74.
[16] Biot MA. General theory of three-dimensional consolidation. J Appl Phys 1941; 12: 155-164.
[17] Karalis TK. Integrated effects on the shrinkage stresses from the water loss in soft cohesive soils. Int J Engineering Sci 2003; 41: 371-385.
[18] Perrier E, Mullon C, Rieu M, de Marsily G. Computer construction of fractal soils structures: Simulation of their hydraulic and shrinkage properties. Water Resour Res 1995; 31: 2927-2943.
[19] Rieu M, Sposito G. Fractal fragmentation, soil porosity, and soil water properties. I. Theory. Soil Sci Soc Am J 1991; 55: 1231-1238.
[20] Moran CJ, McBratney AB. A two-dimensional fuzzy random model of soil pore structure. Math Geol 1997; 29: 755-777.
[21] Horgan GW, Young IM. An empirical stochastic model for the geometry of two-dimensional crack growth in soil (with Discussion). Geoderma 2000; 96: 263-276.
[22] Hornig T, Sokolov IM, Blumen A. Patterns and scaling in surface fragmentation processes. Phys Rev E 1996; 54(4): 4293-4298.
[23] Malthe-Sørenssen A, Wallmann T, Feder J, Jøssang T, Meakin P, Hardy HH. Simulation of extensional clay fractures. Phys Rev E 1998; 54: 5548-5564.
[24] Kitsunezaki S. Fracture patterns induced by desiccation in a thin layer. Phys Rev E 1999; 60: 6449-6464.
[25] Shorlin KA, de Bruyn JR. Development and geometry of isotropic and directional shrinkage-crack patterns. Phys Rev E 2000; 61: 6950-6957.
[26] Vogel H-J, Hoffmann H, Leopold A, Roth K. Studies of crack dynamics in clay soil. II. A physically based model for crack formation. Geoderma 2004; 125: 213-223.
[27] Chertkov VY, Ravina I. Networks originating from the multiple cracking of different scales in rocks and swelling soils. Int J Fracture 2004; 128: 263-270.
[28] Chertkov VY. Intersecting-surfaces approach to soil structure. Int Agrophysics 2005; 19: 109-118.
[29] Chertkov VY. Mathematical simulation of soil cloddiness. Int Agrophysics 1995; 9: 197-200
[30] Bronswijk JJB. Prediction of actual cracking and subsidence in clay soils. Soil Science 1989; 148: 87-93.
[31] Bronswijk JJB. Shrinkage geometry of a heavy clay soil at various stresses. Soil Sci Soc Am J 1990; 54: 1500-1502.

[32]   Olsen PA, Haugen LE. A new model of the shrinkage characteristic applied to some Norwegian soils. Geoderma 1998; 83: 67-81.
[33]   Cornelis WM, Corluy J, Medina H, Hartmann R, Van Meirvenne M, Ruiz ME. A simplified parametric model to describe the magnitude and geometry of soil shrinkage. Europ J Soil Sci 2006; 57: 258-268.
[34]   Zhurkov SN, Kuksenko VS, Petrov VA. Physical principles of prediction of mechanical disintegration. Sov Phys Dokl (Engl Transl) 1981; 26: 755-757.
[35]   Hudson JA, Priest SD. Discontinuities and rock mass geometry. Int J Rock Mech Min Sci Geomech Abstr 1979; 16: 339-362.
[36]   Scott GJT, Webster R, Nortcliff S. An analysis of crack pattern in clay soil: Its density and orientation. J Soil Sci 1986; 37: 653-668.
[37]   Chertkov VY, Ravina I. The effect of interaggregate capillary cracks on the hydraulic conductivity of swelling clay soils. Water Resour. Res. 2001; 37: 1245-1256.
[38]   Chertkov VY, Ravina I. The combined effect of interblock and interaggregate capillary cracks on the hydraulic conductivity of swelling clay soils. Water Resour Res 2002; 38(8): 10.1029/2000WR000094.
[39]   Chertkov VY. Some possible interconnections between shrinkage cracking and gilgai. Austral J Soil Res 2005; 43: 67-71.
[40]   Bronswijk JJB. Drying, cracking, and subsidence of a clay soil in a lysimeter. Soil Science 1991; 152(2): 92-99.
[41]         Bronswijk JJB, Evers-Vermeer JJ. Shrinkage of Dutch clay soil aggregates. Neth J Agric Sci 1990; 38: 175-194.
[42]   Chertkov VY. Estimating the aggregate/intraaggregate mass ratio of a shrinking soil. Open Hydrology Journal 2008; 2: 7-14

17**Figure Captions**

**Fig. (1)**. Scheme of the possible soil depth coordinates: (1) $z_I$ coordinate with the fixed point of reference at the *initial* position of the soil surface; and (2) $z_C$ coordinate with the movable point of reference at a *current* position of the soil surface at a moment $t$. $S(z_I=0, t)$ is a current subsidence of the soil surface at a moment $t$.

**Fig. (2)**. The reference shrinkage curve, $\overline{V}(W)$ of type 1 (with the inflection point in the structural shrinkage area) estimated for Bronswijk's soil [40] from the oven-dried specific volume ($\overline{V}_{az}$) of the intraaggregate matrix, maximum swelling water content of the soil, $W_o$ as well as $\rho_s$, $c$, $P_z$, and $K$ (see Table 1; $K=1.107$) based on the approach [8]. $W$ is the soil gravimetric water content. $\overline{V}_a(w)$ is the experimental shrinkage curve of separate soil aggregates from [40]. The auxiliary shrinkage curve of the intraaggregate matrix is assumed to coincide with $\overline{V}_a(w)$. $w$ is the gravimetric water content of the matrix. For the water content $W_z$, $W_n$, $W_s$, $W_o$, $w_z$, $w_n$, and $w_o$ [8] see Notation.

**Fig. (3)**. As in Fig.**2**, but for the reference shrinkage curve, $\overline{V}(W)$ of type 2 (without an inflection point in the structural shrinkage area) and $K=1.068$.

**Fig. (4)**. The shrinkage geometry factor of the upper (sufficiently thin) soil layer, $r_s$ vs. soil water content (in the layer), estimated for Bronswijk's lysimeter experiment [40] using the modified approach under consideration. The solid line - case of the reference shrinkage curve from Fig.**2** ($K=1.107$). The dashed line - case of the reference shrinkage curve from Fig.**3** ($K=1.068$).

**Fig. (5)**. The cumulative volume of cracks in the large lysimeter core from depth $z_m=50$ cm to a current $z$ value (per unit area) estimated using the modified approach under consideration at 82 d drying duration and the reference shrinkage curve from Fig.**3** ($K=1.068$). 1 - contribution of horizontal cracks. 2 - contribution of vertical cracks. 3 - total crack volume.

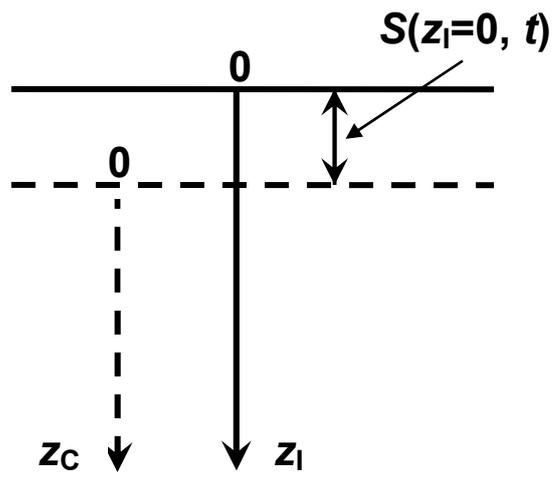

Fig.1

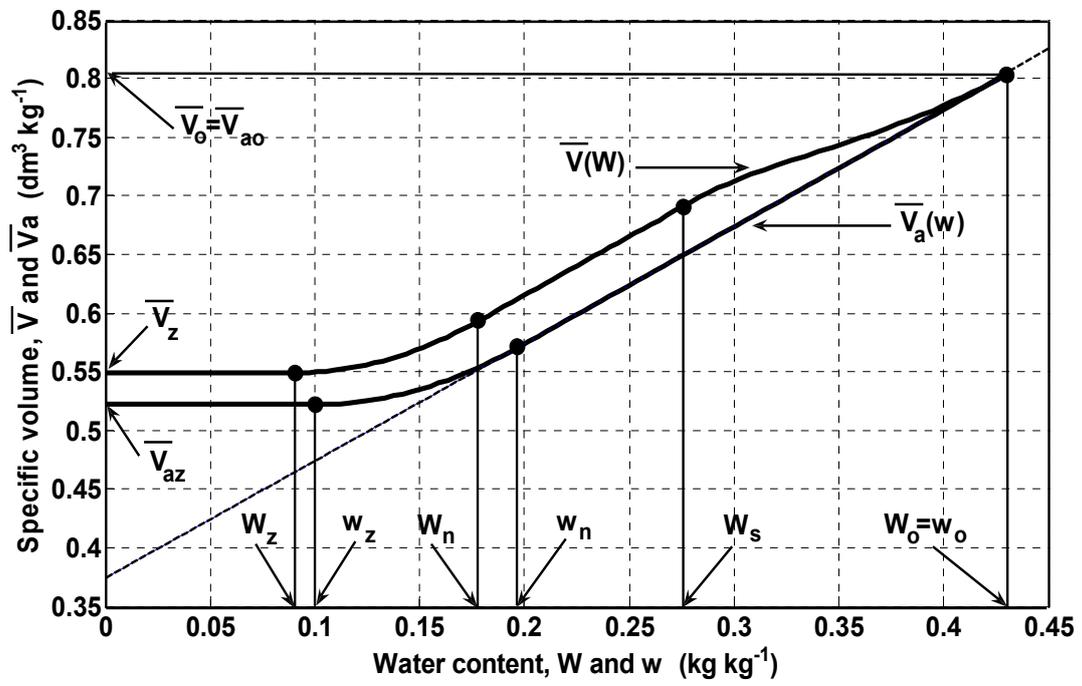

Fig.2

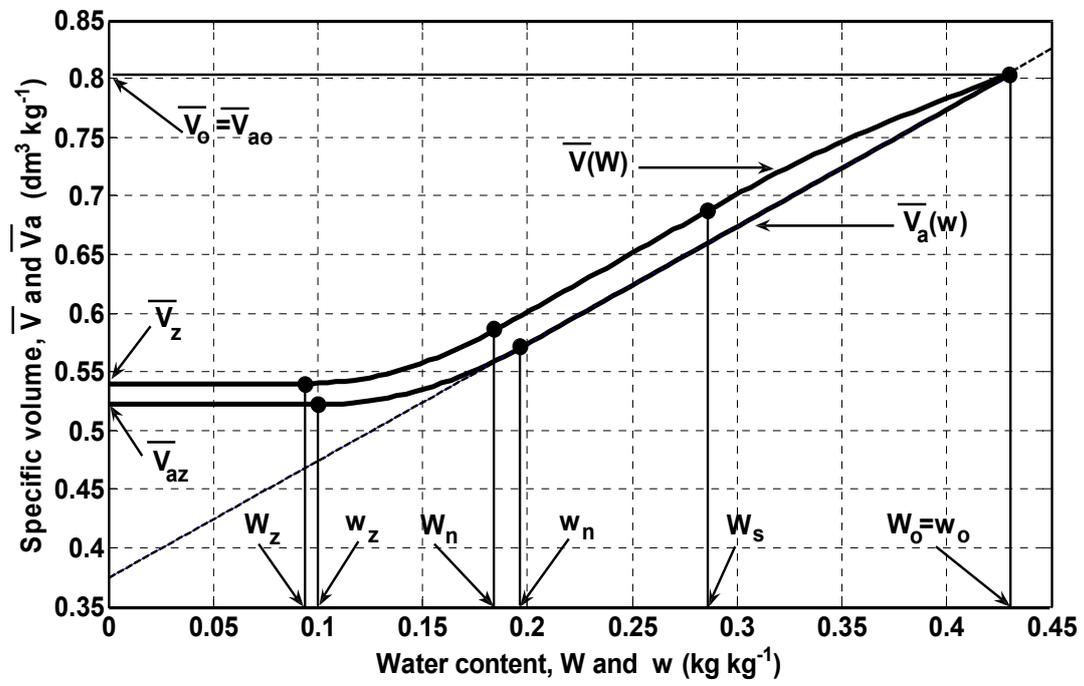

Fig.3

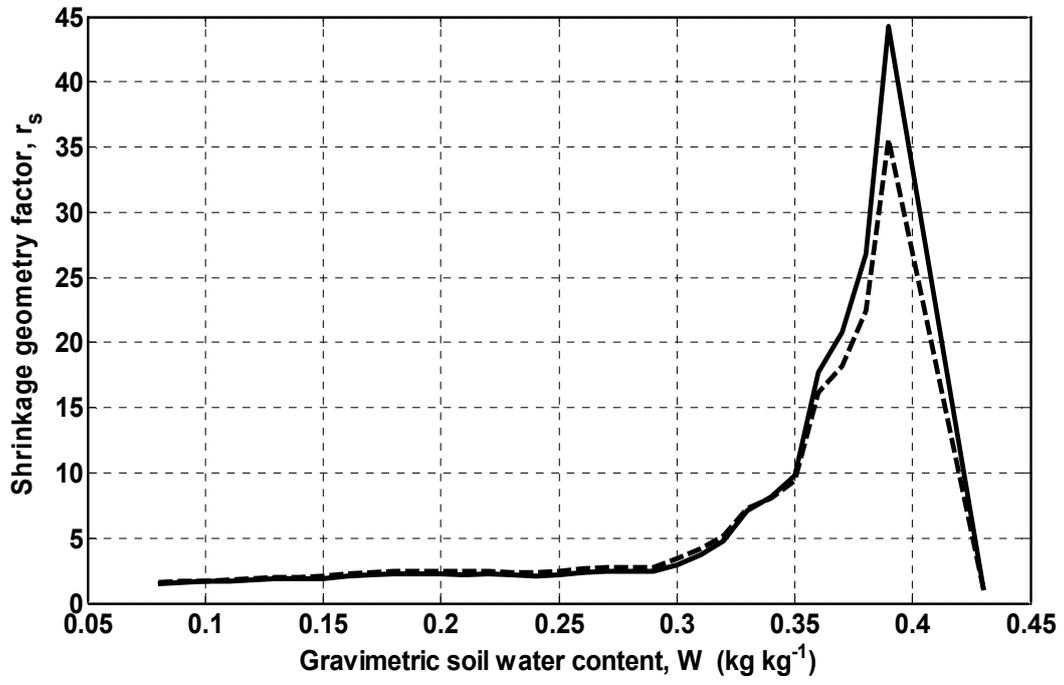

Fig.4

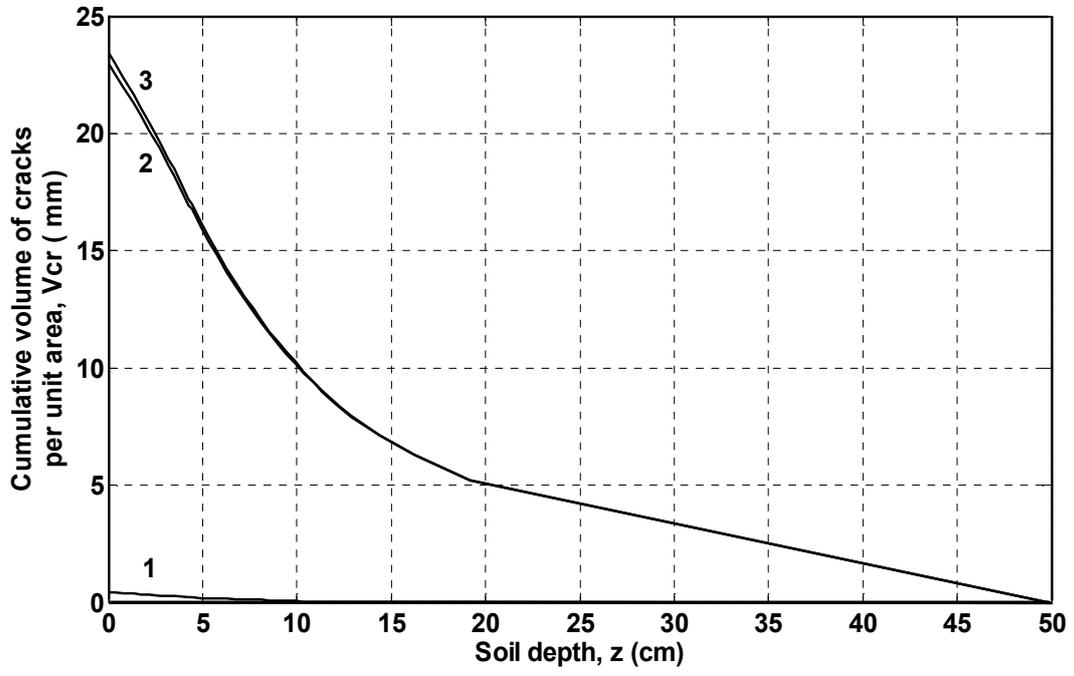

Fig.5

**Table 1. The Values of Parameters that Were Used in the Data Analysis[†]**

| $z_m$ | $z_o$ | $\rho_s$ | $c$ | $W_o$ | $\overline{V}_{az}$ | $P_z$ | K | | $n$ | $T_o$ |
|---|---|---|---|---|---|---|---|---|---|---|
| | | | | | | | **Type 1** | **Type 2** | | |
| **cm** | **cm** | **g cm$^{-3}$** | | **kg kg$^{-1}$** | **dm$^3$ kg$^{-1}$** | | | | | **cm** |
| 50 | 5 | 2.676 | 0.55 | 0.43 | 0.522 | 0 | 1.107 | 1.068 | 0.99 | 7.5 |

[†] $z_m$, maximum crack depth; $z_o$, thickness of an upper soil layer of intensive cracking; $\rho_s$, mean density of soil solids; $c$, soil clay content; $W_o$, maximum swelling point of a soil before shrinkage start; $\overline{V}_{az}$, oven-dried specific volume of intraaggregate matrix per unit mass of the matrix itself; $P_z$, oven-dried structural porosity; $K$, ratio of aggregate solid mass to solid mass of intraaggregate matrix; $n$, coefficient in Eq.(41); $T_o$, initial thickness of the upper 7.5 cm horizontal layer of the lysimeter.

**Table 2. Soil Surface Subsidence: Data and Modeling Results**

| Value Origin | Source | Soil Surface Subsidence at Drying Duration | | |
|---|---|---|---|---|
| | | 33 d | 39 d | 82 d |
| | | mm | | |
| Experiment | Bronswijk [40] | 6.3 | 7.0 | 12.4 |
| Model | This Work | 5.30 | 7.80 | 13.39 |

**Table 3. Cumulative Crack Volume (per unit area): Data and Modeling Results**

| Value Origin | Source | Crack Type | Cumulative Crack Volume (per unit area) at Drying Duration | | |
|---|---|---|---|---|---|
| | | | 33 days | 39 days | 82 days |
| | | | mm | | |
| Experiment | Bronswijk [40] | Vertical and Horizontal | -- | -- | 23.1 |
| Model | Bronswijk [40] | Vertical | 11.7 | 13.2 | 21.7 |
| Model | Chertkov and Ravina [1] | Vertical | 10.2 | 12.5 | 13.7 |
| Model | This Work (RSC[†] from Fig.2; $K$=1.107) | Vertical | 17.12 | 19.22 | 22.70 |
| | | Horizontal | 0.15 | 0.28 | 0.37 |
| | | Vertical and Horizontal | 17.27 | 19.50 | 23.07 |
| Model | This Work (RSC[†] from Fig.3; $K$=1.068) | Vertical | 16.48 | 18.90 | 22.61 |
| | | Horizontal | 0.20 | 0.37 | 0.46 |
| | | Vertical and Horizontal | 16.68 | 19.27 | 23.07 |

[†]RSC, reference shrinkage curve